\documentclass[12pt]{article}
\usepackage{amssymb}
\usepackage{amsmath}
\usepackage{amsthm}
\begin{document}
\font\frak=eufm10 scaled\magstep1
\font\fak=eufm10 scaled\magstep2
\font\fk=eufm10 scaled\magstep3
\font\black=msbm10 scaled\magstep1
\font\bigblack=msbm10 scaled\magstep 2
\font\bbigblack=msbm10 scaled\magstep3
\font\scriptfrak=eufm10
\font\tenfrak=eufm10
\font\tenblack=msbm10

%A continuacion definimos los comandos para utilizar los
%fuentes en modo matematico

%Operadores especiales, abrev. matematicasxs
\def\biggoth #1{\hbox{{\fak #1}}}
\def\bbiggoth #1{\hbox{{\fk #1}}}
\def\sp #1{{{\cal #1}}}
\def\goth #1{\hbox{{\frak #1}}}
\def\scriptgoth #1{\hbox{{\scriptfrak #1}}}
\def\smallgoth #1{\hbox{{\tenfrak #1}}}
\def\smallfield #1{\hbox{{\tenblack #1}}}
\def\field #1{\hbox{{\black #1}}}
\def\bigfield #1{\hbox{{\bigblack #1}}}
\def\bbigfield #1{\hbox{{\bbigblack #1}}}
\def\Bbb #1{\hbox{{\black #1}}}
\def\v #1{\vert #1\vert}             %Para denotar elgrado de #1
\def\ord#1{\vert #1\vert}
\def\m #1 #2{(-1)^{{\v #1} {\v #2}}} %Para denotar el signo (-1)^...
\def\lie #1{{\sp L_{\!#1}}}               %%Lie derivative
\def\pd#1#2{\frac{\partial#1}{\partial#2}}
\def\pois#1#2{\{#1,#2\}} %  un parntesis de Poisson {f,g}
\def\set#1{\{\,#1\,\}}             %  notacin para conjuntos
\def\<#1>{\langle#1\rangle}        %  una forma bilineal <x,a>
\def\>#1{{\bf #1}}                %  notacin para vectores
\def\f(#1,#2){\frac{#1}{#2}}
\def\cociente #1#2{\frac{#1}{#2}}
\def\braket#1#2{\langle#1\mathbin\vert#2\rangle} %% <w|z>
\def\brakt#1#2{\langle#1\mathbin,#2\rangle}           %% <w,z>
\def\dd#1{\frac{\partial}{\partial#1}}
\def\bra #1{{\langle #1 |}}
\def\ket #1{{| #1 \rangle }}
\def\ddt#1{\frac{d #1}{dt}}
\def\dt2#1{\frac{d^2 #1}{dt^2}}
\def\matriz#1#2{\left( \begin{array}{#1} #2 \end{array}\right) }
\def\Eq#1{{\begin{equation} #1 \end{equation}}}
\def\ch{\cosh}
\def\sh{{\sinh}}

%%Abreviaturas de simbolos
\def\bw{{\bigwedge}}      %%from Marmo diff geom
\def\hut{{\scriptstyle \land}}            %%from Marmo diff geom
\def\dg{{\goth g^*}}                                                                                                            %%dual of the Lie algebra
\def\Cdg{{C^\infty(\goth g^*)}}
\def\poi{\{\:,\}}                           %parntesis de Poisson {,}
\def\qw{\hat\omega}                %  omega con sombrero
\def\FL{{\sp F}L}                 %  abreviatura para Transf. Legendre
\def\hFL{\widehat{{\sp F}L}}      %  abreviat. para Transf. Legendreext.
\def\XHMw{\goth X_H(M,\omega)}
\def\XLHMw{\goth X_{LH}(M,\omega)}
\def\ea{\varepsilon_a}
\def\ep{\varepsilon}
\def\mitad{\frac{1}{2}}
\def\x{\times}
\def\cinf{C^\infty}
\def\forms{\bigwedge}                 %  formas
\def\onda{\tilde}
\def\orb{{\sp O}}

%%letras griegas
\def\a{\alpha}
\def\d{\delta}
\def\g{{\gamma }}                  %  gama
\def\G{{\Gamma}}	
\def\La{\Lambda}                   %  lambda
\def\la{\lambda}                   %  Lambda
\def\w{\omega}                     %  una forma simplectica
\def\W{{\Omega}}                   %  Omega
\def\ltimes{\bowtie}

%%letras cal, Bbb, goticas, etc.
\def\roc{{\tilde{\cal R}}}                       %%from Marmo diff geom
\def\cl{{\cal L}}                               %%from Marmo diff geom
\def\V{{\sp V}}                                 %espacio de velocidades
\def\F{{\sp F}}
\def\cv{{{\goth X}}}                    %  un campo vectorial
\def\LG{\goth g}
\def\LH{\goth h}
\def\X{{{\goth X}}}                     %  un campovectorial
\def\R{{\hbox{{\field R}}}}             %%real numbers (Pepin)
\def\big R{{\hbox{{\bigfield R}}}}
\def\bbig R{{\hbox{{\bbigfield R}}}}
\def\C{{\hbox{{\field C}}}}         %%complex numbers (Pepin)
\def\Z{{\hbox{{\field Z}}}}             %%real numbers (Pepin)
\def\N{{\hbox{{\field N}}}}         %%complex numbers (Pepin)
%\def\small C{{\hbox{{\smallfield C}}}}         %%smallcomplex numbers (Pepin)

%%notaciones rm en modo math
\def\ima{\hbox{{\rm Im}}}                               %%Image of a map
\def\dim{\hbox{{\rm dim}}}        %%several definitions
\def\End{\hbox{{\rm End}}}
\def\Tr{\hbox{{\rm Tr}}}
\def\tr{{\hbox{\rm\small{Tr}}}}                %%Trace
\def\lin{{\hbox{Lin}}}
\def\vol{{\hbox{vol}}}
\def\Hom{{\hbox{Hom}}}
\def\rank{{\hbox{rank}}}
\def\Ad{{\hbox{Ad}}}
\def\ad{{\hbox{ad}}}
\def\CoAd{{\hbox{CoAd}}}
\def\coad{{\hbox{coad}}}
\def\Rea{\hbox{Re}}                     %  parte real
\def\id{{\hbox{id}}}                    %  la identidad
\def\Id{{\hbox{Id}}}
\def\Int{{\hbox{Int}}}
\def\Ext{{\hbox{Ext}}}
\def\Aut{{\hbox{Aut}}}
\def\Card{{\hbox{Card}}}
\def\SODE{{\small{SODE }}}
\newcommand{\bea}{\begin{eqnarray}}
\newcommand{\eea}{\end{eqnarray}}

\def\R{\mathbb{R}}
\def\ba{\begin{eqnarray}}
\def\ea{\end{eqnarray}}
\def\be{\begin{equation}}
\def\ee{\end{equation}}
\def\Eq#1{{\begin{equation} #1 \end{equation}}}
\def\R{\Bbb R}
\def\C{\Bbb C}
\def\Z{\Bbb Z}
\def\a{\alpha}                  % alpha
\def\b{\beta}                   % beta
\def\g{\gamma}                  % gamma
\def\d{\delta}                  % delta
\def\bra#1{\langle#1|}
\def\ket#1{|#1\rangle}
\def\goth #1{\hbox{{\frak #1}}}
\def\<#1>{\langle#1\rangle}
\def\cotg{\mathop{\rm cotg}\nolimits}
\def\Map{\mathop{\rm Map}\nolimits}
\def\wt{\widetilde}
\def\const{\hbox{const}}
\def\grad{\mathop{\rm grad}\nolimits}
\def\Div{\mathop{\rm div}\nolimits}
\def\braket#1#2{\langle#1|#2\rangle}
\def\Erf{\mathop{\rm Erf}\nolimits}
\def\matriz#1#2{\left( \begin{array}{#1} #2 \end{array}\right) }
\def\Eq#1{{\begin{equation} #1 \end{equation}}}
\def\deter#1#2{\left| \begin{array}{#1} #2 \end{array}\right| }
\def\pd#1#2{\frac{\partial#1}{\partial#2}}
\def\til{\tilde}

\def\la#1{\lambda_{#1}}
\def\teet#1#2{\theta [\eta _{#1}] (#2)}
\def\tede#1{\theta [\delta](#1)}
\def\N{{\frak N}}
\def\GR{{\cal G}}
\def\Wei{\wp}

\def\pd#1#2{\frac{\partial#1}{\partial#2}}
                                                %  una derivada parcial
\def\matrdos#1#2#3#4{\left(\begin{matrix}#1 & #2 \cr          %para matrices 2X2
                                 #3 & #4 \cr\end{matrix}\right)}
%Si no se puede utilizar el fichero mssymb, los fuentes AmS TeX se
%pueden cargar a mano (por ejemplo) con las lineas siguientes
%\newfont{\got}{eufm10 scaled\magstep1}
%\newfont{\field}{msym10 scaled\magstep1}

\newtheorem{teor}{Teorema}[section]
\newtheorem{cor}{Corolario}[section]
\newtheorem{prop}{Proposici\'on}[section]
\newtheorem{note}[prop]{Note}
\newtheorem{definicion}{Definici\'on}[section]
\newtheorem{lema}{Lema}[section]
%\newexample{ejem}{Ejemplo}[section]
\theoremstyle{plain}
\newtheorem{theorem}{Theorem}
\newtheorem{corollary}{Corollary}
\newtheorem{proposition}{Proposition}
\newtheorem{definition}{Definition}
\newtheorem{lemma}{Lemma}

\def\Eq#1{{\begin{equation} #1 \end{equation}}}
\def\R{\Bbb R}
\def\C{\Bbb C}
\def\Z{\Bbb Z}
\def\mp#1{\marginpar{#1}}

\def\la#1{\lambda_{#1}}
\def\teet#1#2{\theta [\eta _{#1}] (#2)}
\def\tede#1{\theta [\delta](#1)}
\def\N{{\frak N}}
\def\Wei{\wp}
\def\Hil{{\cal H}}

\font\frak=eufm10 scaled\magstep1

\def\bra#1{\langle#1|}
\def\ket#1{|#1\rangle}
\def\goth #1{\hbox{{\frak #1}}}
\def\<#1>{\langle#1\rangle}
\def\cotg{\mathop{\rm cotg}\nolimits}
\def\cotanh{\mathop{\rm cotanh}\nolimits}
\def\arctanh{\mathop{\rm arctanh}\nolimits}
\def\wt{\widetilde}
\def\const{\hbox{const}}
\def\grad{\mathop{\rm grad}\nolimits}
\def\Div{\mathop{\rm div}\nolimits}
\def\braket#1#2{\langle#1|#2\rangle}
\def\Erf{\mathop{\rm Erf}\nolimits}

\centerline{\Large \bf Lie systems and integrability conditions}
\vskip 0.75cm
\centerline{\Large \bf for $t$-dependent frequency harmonic oscillators}
\vskip 0.75cm
\vskip 0.75cm
\date{}

\centerline{ Jos\'e F. Cari\~nena, Javier de Lucas and Manuel F. Ra\~nada}
\vskip 0.5cm

\centerline{Departamento de  F\'{\i}sica Te\'orica, Universidad de Zaragoza,}
\medskip
\centerline{50009 Zaragoza, Spain.}
\medskip
\centerline{28 May 2009}
\medskip


\vskip 1cm

\begin{abstract}
Time-dependent frequency harmonic oscillators (TDFHO's) are studied through the
theory of Lie systems. We show that they  are related
to a certain kind of equations in the Lie group $SL(2,\R)$. Some integrability conditions 
appear as conditions to be able to transform such equations into simpler ones in a very specific way. 
As a particular application of our results we find 
$t$-dependent constants of the motion for certain one-dimensional TDFHO's. %and the general solution for a two-dimensional TDFHO
  Our
approach provides an unifying framework which allows us to apply our developments
to all Lie systems associated with equations in $SL(2,\mathbb{R})$ and to generalise our methods to study any Lie system. 
\end{abstract}

\section{Introduction and motivation.}

\qquad    The search for solutions of autonomous
integrable systems is based on the existence 
 of a sufficient number of constants of the motion. Unfortunately, the extension of the
 theory to include non-autonomous 
systems is a difficult problem and
 many of the properties of $t$-dependent Hamiltonians still
remain as partially understood.  The situation is even worse for the corresponding quantum case for which 
the $t$-evolution is difficult to find, when possible.

The simplest case is that of a linear system because then there is a linear superposition principle, but the difficulty remains in the
search for 
a fundamental set of particular solutions.

A particular case of this linear system of a high physical interest is the $t$-dependent oscillator, i.e. variable-frequency and  variable-mass
oscillator, which can probably be considered as the most frequently studied of such systems \cite{RaR79}--\cite{GoA93} for its relevance in 
many different problems in Optics and Particle Physics. 
Our aim is to revisit the theory of TDFHO's and some other related ones 
from the perspective of the theory of Lie systems \cite{LS}--\cite{CGM07} but without explicitly using the linearity properties trying to develop the theory in 
the framework of Lie systems and therefore following a process which can be extended to other Lie systems with the same Lie group, as the almost ubiquitous 
in Physics  Riccati equation, the Milne-Pinney equation  or the Ermakov system \cite{Mil30}--\cite{CLR08a}.

Lie systems are a generalisation of linear systems and have received much attention during the last years. Each Lie system has associated a 
Lie group and it can be related to another Lie system  
defined by right-invariant vector fields on the group, this last system being related to all Lie systems with such an associated Lie group \cite{CarRamGra}.

Actually, the explicit integration of the given equations depends very much on the explicit values of the $t$-dependent functions defining the system 
and therefore the interest in establishing such an integrability criteria which turn out to be shared for all Lie systems with the same associated Lie algebra.

Quite often the study of such Lie systems is carried out by means of a transformation leading to a vector field and therefore to an autonomous system.
As an example the standard one-dimensional harmonic oscillator
described by the 
 Hamiltonian 
  $$
	 H_0(Q,P) =\frac 12\, \left( P^2 + Q^2 \right)\,,
  $$
% which gives rise to the evolution
% $$Q(T)=\cos (T)\, Q_0+\sin (T)\, P_0\,,\qquad P(T)= -\sin (T)\, Q_0+\cos (T)\, P_0\,,
% $$
 can be transformed under  a two steps transformation 
 into  the  system described by the $t$-dependent Hamiltonian
  \begin{eqnarray}
	 H_F(t,q,p)=\frac 12\, \left( p^2 + F(t) q^2 \right)\,,\label{H_F}
  \end{eqnarray}
with  $F$ being the $t$-dependent function given by
  $$
   F = \frac{1}{\rho}\left(\frac{1}{\rho^3} - \ddot{\rho}\right)\,,
  $$
and where $\rho$ is an arbitrary positive function, $\rho(t)>0$.

In fact, we first change  the time variable
  $$
    T \mapsto t\,,{\quad} t = {\rho}^2(t) T\,,
  $$
and second, a $t$-dependent (but linear) canonical transformation is carried out
  $$\left\{\begin{array}{rcl}
    q &=& {\rho} Q \,,\\ {\rho} p &=& P + {\rho}\dot{\rho}
    Q,\end{array}\right. \qquad \left\{\begin{array}{rcl} Q&=&q/\rho\,,\\
    P&=&{\rho} p - \dot{\rho} q\,.\end{array}\right.\,
  $$
Then, the Hamiltonian  $H_0$ becomes 
the following $t$-dependent function
\begin{equation}
    I_\rho (t,q,p)=\frac 12 \, \left[ ({\rho} p - \dot{\rho} q)^2 +
\left(\frac{q}{\rho}\right)^2 \right]\,.\label{LRinv}
\end{equation}
Such a function  $I_\rho$ is a ($t$-dependent) constant of the motion for the
$t$-dependent Hamiltonian $H_F$. The existence of such a constant of the motion, usually called the
`Lewis invariant'
\cite{Le67,{LR69}},
has been proved by making use of several alternative
procedures (see e.g. Refs. \cite{RaR79,{PrE80},{GoA93}}).
  Notice that every particular choice for the function
${\rho}(t)$ determines a
$t$-dependent oscillator endowed with an associated $t$-dependent constant
of the motion $ I_\rho(q,p,t)$.
Conversely, suppose  we consider  a $t$-dependent oscillator $H_F$ as in (\ref{H_F})
defined by a given function $F(t)$.
Then, a solution of the auxiliary equation
  $$
   \ddot{\rho} + F(t) {\rho} = {\frac {1} {\rho^3}}
  $$
permits constructing the constant of the motion (\ref{LRinv}).
This auxiliary equation had been considered by Milne \cite{Mil30} and its
general 
solution
 was studied in 1950  by Pinney \cite{Pi50} who proved 
that it belongs to the restricted subfamily of nonlinear differential equations
whose exact solutions can be obtained.
Nevertheless, the Pinney method makes use of two independent solutions of
the associated linear problem.
But this linear equation is just the equation of the
$t$-dependent oscillator and we arrive at the annoying conclusion that
(at least for a general function $F(t)$) we need to solve first another
$t$-dependent
equation of motion and only then we can obtain the explicit form of the
constant of the motion $I_\rho$.
In any case, the above mentioned  two steps procedure shows the existence of a
close relationship between the two equations,
the $t$-independent and the $t$-dependent one of this particular oscillator system. These relations have been explained from the point of view of Lie systems in \cite{CLR08a} where it was also studied why both differential equations
(the Milne--Pinney equation and the $t$-dependent harmonic oscillator) have the same difficulty to be solved exactly.

Integrable systems in the classical 
Arnold--Liouville sense are those  having as many independent first-integrals 
in involution as degrees of freedom.
Such a   system is called super-integrable if, in addition, possesses more
independent first-integrals
 than degrees of freedom.
The free particle, the Kepler problem and the harmonic oscillator
with rationally related  frequencies are three  instances of this very particular
class of systems
(for other super-integrable systems see Refs. \cite{Ev90}--\cite{RaS99}
and references therein).

% The two-dimensional harmonic oscillator described by the Hamiltonian
%   $$
% 	 H_{HO}(q_1,q_2,p_1,p_2)  =  {\frac 1 2}\, (p_1^2 + p_2^2)
%            +  {\frac 1 2}\, ({\omega_1}^2 q_1^2 + {\omega_2}^2 q_2^2)
%   $$
% has two one-degree of freedom energies,
% $I_1=H_1$ and $I_2=H_2$, as  first-integrals in involution.
% If ${\omega_2}\ne{\omega_1}$ the angular momentum is not preserved since the
% potential is not central. Nevertheless, in the very particular case in
% which the quotient of the two frequencies is rational the system has
% a third additional nonlinear first-integral \cite{Ta89}--\cite{CMR02}.
% In geometric terms, the phase space is foliated by tori
% and every integral curve is a curve with constant slope on a torus.
% The slope of the curve is determined by the ratio ${\omega_2}/{\omega_1}$.
%  If this ratio is irrational the corresponding curve is
% dense in the torus. On the other hand, if this ratio is rational the orbit becomes closed and the motion 
% is periodic.

The main objective of this paper is the study of  integrability conditions, in
particular, those for Riccati equations or analogous systems. Such integrability conditions have been extensively studied 
since the investigation made by Liouville \cite{Ka57}. Lots of papers have treated this topic since then \cite{Na99}-\cite{Pr80}. Recently, the theory of Lie systems has been used to
 analyse this subject \cite{CRL07b}--\cite{CLR08b}. These works show that the theory of Lie systems allows us to recover some very well-known results about the integrability of Riccati 
 equations scattered in the literature and to collect them from an unifying perspective. In spite of their apparent simplicity, the methods already developed reduce the problem 
 of integrability of Riccati equations to studying integrability conditions for a particular equation in the Lie group $SL(2,\mathbb{R})$.  Such an equation describes the integral 
 curves for the $t$-dependent vector fields on $SL(2,\mathbb{R})$ made up by a linear combination with $t$-dependent coefficients of right-invariant vector fields with respect to
  the right action of $SL(2,\mathbb{R})$ on itself. Moreover, the theory of Lie systems shows that we can apply these integrability conditions to any other Lie system related to an
   equation in $SL(2,\mathbb{R})$ of the same form. For instance, here we use  results on integrability of Riccati equations to study TDFHO's which are associated with the
    same kind of equations in $SL(2,\mathbb{R})$ as Riccati equations. In this way, we illustrate how the theory of Lie systems provides a way to extend the methods of the theory of 
    integrability of differential equations to integrability of non-autonomous systems in Physics or other Lie systems. Furthermore, the point of view of the theory of Lie systems 
    permits the  application of the integrability conditions of common Lie systems,
    here Riccati equations, to the so-called SODE Lie systems, like TDFHO's, 
which may be understood as Lie systems \cite{CLR08a}.  

The paper is organised as follows. Section 2 is devoted to show that TDFHO's are SODE Lie systems related to a Lie algebra of vector fields isomorphic 
to $\mathfrak{sl}(2,\mathbb{R})$. In Section 3 we
study some transformation properties for these differential equations that turn out to be a
straightforward generalisation of those found for Riccati equation
\cite{CarRam}. Our main result is contained in Section 4 where it is shown that  the theory of Lie systems describes these transformations by
means of a special kind of matrix Riccati equation. In our particular instance, this differential equation is again a Lie system related to a Lie algebra of vector fields isomorphic to
 $\mathfrak{sl}(2,\mathbb{R})\oplus\mathfrak{sl}(2,\mathbb{R})$.
  The integrability conditions for TDFHO's appear as
 conditions to be able to obtain solutions of the differential equation defining the
transformation when we assume certain conditions on the solutions. This
property  is studied in a more detailed way 
in Section 5 where some previously known integrability conditions are recovered
from this new perspective. Some of such an integrability conditions are applied in Sections 6 to
TDFHO's and in particular to Caldirola--Kanai TDFHO's. In Section 7 we develop
a new integrability condition and some new integrable TDFHO's are studied.
 A method to obtain $t$-dependent constants of the motion is shown  in
Section 8. % and 
%integrability conditions  are applied  in Section 9 to a $t$-dependent frequency 
%two-dimensional harmonic oscillator.

\section{TDFHO as a Lie system.}\label{TDFHOLS}

\qquad Any TDFHO can be considered as 
a Lie system related to a Lie algebra of vector fields, known as the Guldberg--Vessiot Lie algebra, isomorphic to $\goth{sl}(2,\R)$. Therefore, we can associate it with an equation in the Lie group $G=SL(2,\R)$, see \cite{CLR08a,CLR08b}. 
In fact,  consider the harmonic oscillator with $t$-dependent frequency and mass described by the Hamiltonian $H\in C^\infty(\R\times{\rm T}^*\mathbb{R})$
$$
H(t,x,p)=\frac 1{2m(t)}p^2+\frac 12F(t)\omega^2x^2\,,
$$
giving rise to the Hamilton equations 
\begin{equation}\label{TDFHO2}
\left\{\begin{aligned}
\dot x&=\frac{\partial H}{\partial p}=\frac{p}{m(t)},\\
\dot p&=-\frac{\partial H}{\partial x}=-F(t)\omega^2x,
\end{aligned}\right.
\end{equation}
which is a non-autonomous system of first-order ordinary differential equations in ${\rm T}^*\R$. Its
solutions are the integral curves for the $t$-dependent vector field 
$$X(t)=\frac{p}{m(t)}\,\pd{}x-F(t)\omega^2x\pd{}p\,.
$$
 Let $X_0, X_1$ and $X_2$ be  the vector fields given by
\begin{equation}\label{VectField}
X_0=p\pd{}{x},\quad X_1=\frac 12\left(x\pd{}{x}-p\pd{}{p}\right),\quad X_2=-x\pd{}{p}\,,
\end{equation}
which close on a Lie algebra with commutation relations
\begin{equation*}
[X_0,X_2]=2X_1,\quad [X_0,X_1]=X_0,\quad [X_1,X_2]=X_2,
\end{equation*}
and therefore isomorphic to $\goth{sl}(2,\R)$.
Then, the $t$-dependent vector field $X$ associated with the system (\ref{TDFHO2})
can be written as a linear combination
\begin{equation}\label{TDFHO4}
 X(t)=F(t)\omega^2X_2+\frac{1}{m(t)}X_0\,,
\end{equation}	
i.e. it is a linear combination with $t$-dependent coefficients 
\begin{equation*}
X(t)=\sum_{\alpha=0}^2b_\alpha(t)X_\alpha,
\end{equation*}
 with $b_0(t)=1/m(t)$, $b_1(t)=0$ and $b_2(t)=F(t)\omega^2$. Hence, $t$-dependent mass and frequency harmonic oscillators are SODE
 Lie systems. Moreover, for constant mass, i.e. $m(t)=1$, we get that TDFHO's are SODE Lie systems. 

Let $V$ be the real Guldberg--Vessiot Lie algebra 
 generated by the complete vector fields
  $\{X_\alpha\,|\,\alpha=0,1,2\}$ and isomorphic to  $\goth{sl}(2,\R)$.
It is  known  that if the vector fields of such a Lie algebra 
  are complete, there is an action
 $\Phi_{HO}$ of  the  connected Lie group $SL(2,\R)$ 
on  the manifold ${\rm T}^*\R$ whose fundamental vector fields are those of
$V$.

Let us choose a basis $\{M_0,M_1,M_2\}$ of $\goth{sl}(2,\R)$  such that
the fundamental vector fields corresponding to the $\{M_\alpha\,|\,\alpha=0,1,2\}$ are
the respective  $\{X_\alpha\,|\,\alpha=0,1,2\}$. For example, if we consider that  $\goth{sl}(2,\R)$  can be identified  with the set of 
traceless $2\times 2$ matrices, we can choose the basis 
\begin{equation}\label{Bas}
M_0=\left(\begin{array}{cc}
0&-1\\
0&0
\end{array}\right),\quad
M_1=\frac 12\left(\begin{array}{cc}
-1&0\\
0&1
\end{array}\right),\quad
M_2=\left(\begin{array}{cc}
0&0\\
1&0
\end{array}\right),
\end{equation}
which close the same commutation relations as the
$X_\alpha$. 
Thus, the linear map such that the image of  $M_\alpha$ is $ X_\alpha$ is
an isomorphism 
of Lie algebras and we can relate (\ref{TDFHO2}) to an equation in $SL(2,\R)$ given by
\begin{equation}\label{eLA}
R_{A^{-1}*A}\dot A=-\sum_{\alpha=0}^2b_\alpha(t)M_\alpha,
\end{equation}
with the initial condition $A(0)=I$ (see \cite{CLR08b} for the reason of the minus sign). Thus, if $A(t)$ is the solution  of  (\ref{eLA}) and we denote  $\xi=(x,p)\in{\rm T}^*\mathbb{R}$, then 
the solution of (\ref{TDFHO2}) starting from $\xi(0)$ is $\xi(t)=\Phi_{HO}(A(t),\xi(0))$ (see e.g. \cite{CarRamGra}).

For any $t$-dependent harmonic oscillator, it can be verified that the left action 
$\Phi_{HO}$ of $SL(2,\R)$ on ${\rm T}^*\R\approx \R^2$ associating the elements of the basis (\ref{Bas}) with their fundamental vector fields (\ref{VectField}) is the linear action given by
\begin{equation*} 
\Phi_{HO}\left(A,\left(\begin{array}{c}
x\\p\end{array}\right)\right)=
\left(\begin{array}{cc}
\alpha&\beta\\
\gamma&\delta
\end{array}
\right)\left(\begin{array}{c}
x\\p\end{array}\right) {\rm with }\quad 
A\equiv\left(\begin{array}{cc}
\alpha&\beta\\
\gamma&\delta
\end{array}
\right)\in SL(2,\R).
\end{equation*}
In this way, the one-parameter subgroup 
$\exp(-t\,M_\alpha)$ acts on ${\rm T}^*\R$ with 
infinitesimal generator $X_\alpha$.

In summary, the system (\ref{TDFHO2}) is a Lie system in ${\rm
  T}^*\R$ related
to an equation in $SL(2,\R)$ and the solutions of the latter  equation
 allow us to obtain the solutions of (\ref{TDFHO2}) in terms of the
 initial condition by means of the action $\Phi_{HO}$. We can treat in a similar way
 other Lie systems with an isomorphic Guldberg--Vessiot Lie algebra, as for
 instance the Milne-Pinney equation, but the action is not linear anymore.
\section{Transformations of Lie systems in $SL(2,{\mathbb{R}})$.}\label{TL}

\qquad We first recall an important property of Lie systems associated with a Lie group $G$ that in the
particular case of TDFHO's plays a very relevant r\^ole for
establishing  integrability
criteria: the group $\mathcal{G}\equiv{\rm Map}(\R,G)$
of curves in the Lie group $G$ , here  $SL(2, {\R})$, acts on the set  of the
Lie systems with such a group, see \cite{CarRam,CRL07b,CLR08b}.

More explicitly, in the case we are considering each $t$-dependent frequency harmonic oscillator (\ref{TDFHO2}) can be considered as a
curve in ${\R}^3$ through the identification with the Lie algebra
$\goth{sl}(2,\R)$ corresponding to the choice of its basis  $\{M_1,M_2,M_3\}$,  i.e. the curve in $\R^3$ is defined by the coordinates
of the curve $M(t)$ in $\goth{sl}(2,\R)$
with respect to such  basis
 associated with  the  $t$-dependent vector 
field  defining the TDFHO. Actually, 
we can
transform every curve $\xi(t)$ in ${\rm T}^*\R$,
by an element $\bar A(t)$ of $\mathcal{G}$ as follows:
\begin{eqnarray} {\rm If}  \ \bar A(t)=\!\matriz{cc}
  {{\bar{\a}(t)}&{\bar{\b}(t)}\\{\bar{\g}(t)}&{\bar{\d}(t)}}\in{\cal G},\qquad 
\Theta(\bar A,\xi)(t)=\!
\left(\begin{aligned}
{\bar{\a}(t) x(t)+\bar{\b}(t) p(t)}\\
{\bar{\g}(t) x(t)+\bar{\d}(t)p(t)}
\end{aligned}\right)\!.\label{acciondecurva}
\end{eqnarray}
It can be checked \cite{CarRam} that the curve $\bar A(t)$ transforms the TDFHO (\ref{TDFHO2})
into an analogous 
 TDFHO with new coefficients $b'_0,b'_1, b'_2$ given by
\bea\label{trans}
b'_2&=&{\bar\d}^2\,b_2-\bar\d\bar\g\,b_1+{\bar\g}^2\,b_0+\bar\g {\dot{\bar\d}}-\bar\d \dot{\bar\g}\ ,
\nonumber \\
b'_1&=&-2\,\bar\b\bar\d\,b_2+(\bar\a\bar\d+\bar\b\bar\g)\,b_1-2\,\bar\a\bar\g\,b_0
       +\d \dot{\bar\a}-\bar\a \dot{\bar\d}+\bar\b \dot{\bar\g}-\bar\g \dot{\bar\b}\ ,   \\
b'_0&=&{\bar\b}^2\,b_2-\bar\a\bar\b\,b_1+{\bar\a}^2\,b_0+\bar\a\dot{\bar\b}-\bar\b\dot{\bar\a} \ .
 \nonumber
\eea

In fact, this expression defines an affine action (see e.g. \cite{LM87} for the general definition of
this concept) of the group
 ${\GR}$ on the set of
TDFHO's \cite{CarRam}.
This means that in order  to transform the coefficients of a TDFHO by means of
two transformations of the above type, firstly through $A_1$ and then by means of $A_2$, 
it suffices to do the transformation defined by
the product  element $A_2\,A_1$ of
${\GR}$.

The result of this action of $\mathcal{G}$ can also be studied from the point of
view
of the equations in $SL(2,\mathbb{R})$. First, $\mathcal{G}$  acts on the left on the
set of curves in $SL(2, \R)$ by left translations, i.e. a curve $\bar A(t)$
transforms the curve $A(t)$  into a new one  $A'(t)=\bar A(t) A(t)$. Therefore, if
 $A(t)$ is a solution of (\ref{eLA}), characterised by a curve $M(t)\in T_ISL(2,\R)$,
then the new curve satisfies a new equation like (\ref{eLA}) but  with a
different right-hand side,
 $M'(t)$, and thus it corresponds to a new equation in
$SL(2,\mathbb{R})$ associated with a new TDFHO. Of course, $A'(0)=\bar A(0)A(0)$,
and if we want $A'(0)={\rm Id}$ we have to impose the additional condition 
$\bar A(0)={\rm Id}$. In this way
$\mathcal{G}$
 acts on the set of curves in $T_ISL(2,\R)\simeq\goth{sl}(2,\R)$. It can be shown that
 the relation between both
curves $M(t)$ and $M'(t)$ in $T_{I}SL(2,\mathbb{R})$ is given by \cite{CarRamGra}
\begin{equation}
M'(t)=-\sum_{\alpha=0}^2b'_\alpha(t)M_\alpha=\bar A(t)M(t)\bar A^{-1}(t)+\dot{\bar{A}}(t)\bar A^{-1}(t)
\,. \label{newTDHO}
\end{equation}

Summarising, it has been  shown that it is possible to associate, in a
one-to-one way, TDFHO's with equations in
the Lie group $SL(2,\mathbb{R})$ and to define a group $\mathcal{G}$ of transformations on
the set of such TDFHO's induced by the natural group action of $SL(2,\mathbb{R})$.

\section{Lie structure of an equation of transformation of Lie systems.}
\qquad Our aim in this Section is to  construct, given a  pair of equations in $SL(2,\R)$ characterised by two curves $M(t), M'(t)\subset T_ISL(2,{\mathbb{R}})$,  
a system of differential equations which  is a Lie system and whose integral curves
 relate both systems. 
This system of differential equations is used   in next Sections 
to extend the developments of \cite{CarRamGra, CRL07b} to a broader set of cases.

Now, let us consider (\ref{newTDHO}) as a system of first-order ordinary differential equations in the
coefficients of the
 curve in $SL(2,\R)$,
$$
\bar A(t)=
\matriz{cc}{
\alpha(t) &\beta(t)\cr
\gamma(t)& \delta(t)}\,,
$$
relating both systems.

Rewrite (\ref{newTDHO}) in a more useful way by multiplying on the right by $\bar A(t)$ to obtain
\begin{equation}
\dot{\bar{A}}(t)=M'(t)\bar A(t)-\bar A(t)M(t),\,\label{MReq}
\end{equation}
that is, a special kind of matrix Riccati equation, see \cite{AHW82}. Even if it is known that matrix Riccati equations
are Lie systems, nevertheless, we consider convenient to prove this fact in
detail 
in order to show later on how to generalise our methods.

In terms of the coefficients of $ \bar A(t)$, this differential equation  is
\begin{equation}\label{FS}
\matriz{c}{
\dot\alpha\\
\dot\beta\\
\dot\gamma\\
\dot\delta}
=
\matriz{cccc}{
\frac{b'_1-b_1}{2}&b_2 &b'_0&0\\
-b_0& \frac{b'_1+b_1}{2}&0 &b'_0\\
-b'_2&0 &-\frac{b'_1+b_1}{2}& b_2\\
0&-b_2' &-b_0& -\frac{b'_1-b_1}{2}}
\matriz{c}{
\alpha\\
\beta\\
\gamma\\
\delta},
\end{equation}
with
$$
	M(t)=-\sum_{\alpha=0}^2b_\alpha(t)M_\alpha\,,\quad M'(t)=-\sum_{\alpha=0}^2b'_\alpha(t)M_\alpha\,.
$$

It seems that in order to consider curves in $SL(2,\R)$ 
 we should also  to impose that at any time
 $\alpha\delta-\beta\gamma=1$.
 Nevertheless, it is
 shown later on that we can drop such a restriction for the time being because it is
automatically implemented by imposing a restriction on the initial
 conditions. 
Therefore we can deal with  $\delta$ in the
 preceding system of
 differential equations (\ref{FS}) as being independent  of the other variables. This linear
 system of differential equations can be understood as a Lie system 
 associated with a Lie algebra of vector fields isomorphic to $\mathfrak{gl}(4,\mathbb{R})$ but it may be seen in an alternative and
 more interesting
way  as a Lie system in one of its Lie subalgebras. In fact, let us consider the set of vector fields

{\footnotesize
\begin{equation*}
\begin{aligned}
&N_0=-\alpha\pd{}{\beta}-\gamma\pd{}{\delta}, &\, &N'_0=\gamma\pd{}{\alpha}+\delta\pd{}{\beta},\cr
&N_1=\frac
12\left(-\alpha\pd{}{\alpha}+\beta\pd{}{\beta}-\gamma\pd{}{\gamma}+\delta\pd{}{\delta}\right),
&\,&
N'_1=\frac 12\left(\alpha\pd{}{\alpha}+\beta\pd{}{\beta}-\gamma\pd{}{\gamma}-\delta\pd{}{\delta}\right),\cr
&N_2=\beta\pd{}{\alpha}+\delta\pd{}{\gamma},&\,&N'_2=-\alpha\pd{}{\gamma}-\beta\pd{}{\delta},\nonumber
\end{aligned}
\end{equation*}}
for which  the non-null commutation relations among them are given by
\begin{eqnarray*}
&&\left[ N_0,N_1\right]=N_0, \qquad [N_0,N_2]=2  N_1, \qquad [N_1,N_2]=N_2,\cr
&&[N'_0,N'_1]=N'_0, \qquad [N'_0, N'_2]=2
  N'_1,\qquad [N'_1,  N'_2]=N'_2\,.\nonumber
\end{eqnarray*}

If we  denote  $x\equiv\left(\alpha,\beta,\gamma,\delta\right)\in \R^4$ 
the column vector of the four elements of the matrix 
$$\bar A= \matriz{cc}{\alpha&\beta\\ \gamma&\delta},
$$
then
 (\ref{FS}) is a system of   differential equations in $\R^4$ whose
 solutions give us the integral curves $x(t)$ for the $ t$-dependent vector field 
\begin{equation}
N(t)=\sum_{\alpha=0}^2(b_\alpha(t)N_\alpha+b_\alpha'(t)N'_\alpha).\label{TRR4}
\end{equation}

Note that $[N_i,N'_j]=0$, for $i,j=0,1,2$, and therefore
the linear differential equation (\ref{FS}) is a Lie system in $\R^4$ 
associated with a Lie algebra of vector fields isomorphic to
$\LG\equiv\goth{sl}(2,\R)\oplus\goth{sl}(2,\R)$, i.e.
this Lie
algebra decomposes into a direct sum of two Lie algebras isomorphic to
$\goth{sl}(2,\R)$ generated by $\{N_0,N_1,N_2\}$ and 
 by $\{N'_0,N'_1,N'_2\}$, respectively.

The vector fields of the set $\{N_0, N_1, N_2, N'_0,N'_1,N'_2\}$
generate an integrable  distribution of rank
three in almost any point of $\R^4$. Consequently, there exists, at least locally, a
$t$-independent constant of the 
 motion which turns out to be $\det \bar
 A=\alpha\delta-\beta\gamma$. Such a determinant  is a first-integral for all
 the vector fields of the module of vector fields generated by
  $\{N_0,N_1, N_2, N'_0,N'_1,N'_2\}$
and, if we have a solution of  (\ref{FS})
with an initial condition $\det \bar A(0)=\alpha(0)\delta(0)-\beta(0)\gamma(0)=1$,
then $ \det \bar A(t)=1$  at any time $t$ and the solution can be understood
as a curve in $SL(2,\R)$. Therefore, we have found that
the curves in $SL(2,\R)$
relating the  two different curves associated with two TDFHO's can be
described by the curves $A(t)$ that are defined by the integral curves of (\ref{TRR4}) with
$\det \bar A(0)=1$, 
and viceversa.

\begin{theorem}\label{THLS} The curves in $SL(2,\R)$  relating two equations in the Lie group
  $SL(2,\R)
$ characterised by the curves in $T_ISL(2,\mathbb{R})$
$$
M'(t)=-\sum_{\alpha=0}^2b'_\alpha(t)M_\alpha\,,
\qquad M(t)=-\sum_{\alpha=0}^2b_\alpha(t)M_\alpha,$$
 are given by the integral curves  of the $ t$-dependent vector field 
\begin{equation*}
N(t)=\sum_{\alpha=0}^2\left(b_\alpha(t)N_\alpha+b_\alpha'(t)N'_\alpha\right)\,,
\end{equation*}
such that $\det \bar A(0)=1$. This system is a Lie system associated with a non-solvable Lie algebra of vector fields isomorphic to
$\goth{sl}(2,\R)\oplus\goth{sl}(2,\R)$.
\end{theorem}
\begin{corollary} \label{CorCur} Given two TDFHO's associated with the
  curves $M'(t)$ and $M(t)$ in
$\goth{sl}(2,\R)$, there  always exists 
a curve in $SL(2,\R)$ relating both systems.
\end{corollary}
\begin{proof}
It is a direct consequence of the existence and unicity theorem of solutions of differential equations.
\end{proof}

We must remark that even if we know that given two equations in the Lie group $SL(2,\R)$ there
 always exists a transformation relating both, in order to find such a
curve  we need to solve  the system of
differential
equations providing the integral curves of (\ref{TRR4}).  This 
is the solution of a system of differential
equations that is a Lie system related to a non-solvable Lie
algebra in general. Hence,
 it is
not  easy to find its solutions, i.e.  it may not be integrable by
quadratures.

The result  of Theorem  \ref{THLS},  i.e.  the system of differential
equations describing the transformations of Lie systems in $SL(2,\mathbb{R})$
is a matrix Riccati equation associated, as a Lie system, with a Lie algebra
$\mathfrak{sl}(2,\mathbb{R})\oplus\mathfrak{sl}(2,\mathbb{R})$, suggests us a method to find some sufficiency conditions
for  integrability of the TDFHO's to be explained in next section.

\section{A review on some known integrability conditions.}\label{DIC}

\qquad Let us first remark that the Lie equations in $SL(2,\R)$ defined by a constant curve,
$M(t)= -\sum_{\alpha} c_\alpha M_\alpha$,
are integrable and, consequently, the same happens with  
curves of the form $M(t)= -D(t)\left(\sum_{\alpha} c_\alpha M_\alpha\right)$, where $D$ is an arbitrary but constant sign function,
because a $t$-reparametrisation by the function $D(t)$ reduces the problem to
the previous one.
In this Section we study some cases when it is possible to find 
curves $\bar A(t)$ in $SL(2,\R)$ relating a
given equation in $SL(2,\R)$ characterised by a curve $M(t)$ to an
equation in $SL(2,\R)$ characterised by a curve of the type $M'(t)=-D(t)(c_0M_0+c_1M_1+c_2M_2)$. This happens if 
equation (\ref{FS}) is easy to solve; the transformation
establishing the relation to such a TDFHO allows us to find
by quadratures  the solution of the given equation. We first restrict ourselves to
study cases in which 
the curve $\bar A(t)$ lies in a one-parameter subset of $SL(2,\R)$. The
results 
we show next are a direct translation to the framework of TDFHO's
of the results given in \cite{CRL07b} for Riccati equations, for a detailed proof see \cite{CL09Ricc}.
 
\begin{theorem}\label{TU} The necessary and sufficient conditions
for the existence of a  transformation
\begin{equation*}
\xi'=\Phi_{HO}(\bar A_0(t),\xi),\qquad \xi=\left(\begin{array}{c}x\cr p\cr\end{array}\right),
\end{equation*}
with
\begin{equation}
\bar A_0(t)=\left(\begin{array}{cc}
\alpha(t)&0\\
0&\alpha^{-1}(t)
\end{array}
\right),\qquad \alpha(t)>0\,,\label{gcero}
\end{equation}
 relating the TDFHO associated with the $t$-dependent vector field 
\begin{equation}
X(t)=b_0(t)X_0+b_1(t)X_1+b_2(t)X_2, \label{binX}
\end{equation}
where $b_0(t)b_2(t)$ has a constant sign,  $b_0(t)b_2(t)\ne 0$, to another integrable one given by
\begin{equation}
X'(t)=D(t)(c_0X_0+c_1X_1+c_2X_2)\,,\label{integrableuno}
\end{equation}
where $c_i$, with $i=0,1,2$, are real numbers such that $c_0c_2\neq 0$,
are
\begin{equation*}
D^2(t)c_0c_2=b_0(t)b_2(t),\qquad {b_1(t)+\frac{1}{2}\left(\frac{\dot
      b_2(t)}{b_2(t)}-\frac{\dot b_0(t)}{b_0(t)}\right)}=c_1\sqrt{
\frac{b_0(t)b_2(t)}{c_0c_2}}.
\end{equation*}

Then, the  transformation is uniquely defined by
\begin{equation*}
\bar A_0(t)=\left(\begin{array}{cc}\left(\frac{b_2(t)c_0}{b_0(t)c_2}\right)^{1/4}&0\\
0&\left(\frac{b_2(t)c_0}{b_0(t)c_2}\right)^{-1/4} \end{array}\right)\,.
\end{equation*}
\end{theorem}
Note that one coefficient, either $c_0$ or $c_2$, can be reabsorbed with a
redefinition of the function $D$.
As a straightforward application of the preceding Theorem, which can be found in a
similar form to that of  \cite{CRL07b}, we obtain the following corollaries:
\begin{corollary}\label{CTU}
The TDFHO (\ref{TDFHO2}) with $b_0(t)b_2(t)\neq 0$ is integrable by a $t$-dependent change of variables 
\begin{equation*}
\xi'=\Phi_{HO}(\bar A_0(t),\xi),
\end{equation*}
with $\bar A_0$ given by (\ref{gcero}),
if and only 
\begin{equation}
\sqrt{\frac{c_0c_2}{b_0(t)b_2(t)}}\left[b_1(t)+\frac{1}{2}\left(\frac{\dot b_2(t)}{b_2(t)}-\frac{\dot b_0(t)}{b_0(t)}\right)\right]=c_1\,,\label{Kcond}
\end{equation}
for certain real constants $c_0, c_1$ and $c_2$.
\end{corollary}
In this case 
$$D(t)=\sqrt{\frac{b_0(t)b_2(t)}{c_0c_2}},$$ and the new system is 
\begin{equation}
\frac{d\xi'}{dt}=D(t)\matriz{cc}{c_1/2&c_0\\-c_2&-c_1/2}\,\xi'\,.\label{Kevol}
\end{equation}

\begin{corollary}\label{C2TU} Given an integrable TDFHO characterised by a  $t$-dependent vector field (\ref{integrableuno}), 
the set of TDFHO's which can be  obtained through a $t$-dependent 
transformation \begin{equation*}
\xi'=\Phi_{HO}(\bar A_0(t),\xi),
\end{equation*}
with $\bar A_0$ given by (\ref{gcero}), are
those of the form
\begin{equation}
X(t)=b_0(t)X_0+\left( \frac{\dot b_0(t)}{b_0(t)}-\frac{\dot D(t)}{D(t)}+c_1D(t)\right)X_1+\frac{D^2(t)c_0c_2}{b_0(t)}X_2\,.\label{Xdetbcero}
\end{equation}
Thus $\bar A_0(t)$ reads
\begin{equation*}
\bar A_0(t)=\left(\begin{array}{cc}\left(\frac{b_2(t)c_0}{b_0(t)c_2}\right)^{1/4}&0\\
0&\left(\frac{b_2(t)c_0}{b_0(t)c_2}\right)^{-1/4} \end{array}\right)\,.
\end{equation*}
\end{corollary}

Therefore, starting from an integrable system we can find a family
(\ref{Xdetbcero}) of
solvable TDFHO systems whose coefficients are parametrised by $b_0(t)$. 
Given a 
 TDFHO it is easy to check whether it belongs to such a family 
 and can be easily integrated.

The integrability conditions we have described here arise as  
requirements on the initial $t$-dependent functions $b_\alpha$ that allow to solve exactly the initial TDFHO by a $t$-dependent transformation of the form
\begin{equation*}
\xi'=\Phi_{HO}(\exp(\Psi(t)v),\xi),
\end{equation*}
with some $v\in\goth{sl}(2,\R)$ and  $\Psi(t)$, in such a way that the initial TDFHO system
(\ref{TDFHO2}) in the variable $\xi$ transforms into another one in the
variable $\xi'$ that is  a Lie system with a  Guldberg--Vessiot Lie algebra
of vector fields isomorphic to an appropriate 
 Lie subalgebra of $\mathfrak{sl}(2,\R)$ in such a way that 
 the equation in $\xi'$
can be integrated by quadratures and the equation in $\xi$ is solvable too.

\section{Applications of integrability conditions.}

\qquad As a first application we show that the usual approach to the
 solution of the classical Caldirola--Kanai
Hamiltonian \cite{Cal41, Ka48} (the solution of the quantum case can be
worked in a similar way and it will be explained in a forthcoming paper) can be explained through our method. We
also apply our approach to  TDFHO's. In particular, we study the case of 
a frequency 
 $$\Omega(t)=\frac{\omega^2}{(K-c_1t\omega)^2}\,.$$

The Hamiltonian of the  $t$-dependent harmonic oscillator is 
\begin{equation}
H(t,x,p)=\frac 12\,\frac {p^2}{m(t)}+\frac 12\, m(t)\omega^2(t)x^2\,.\label{TDHOH}
\end{equation}
% which gives rise to  the following 
% Hamilton equations
% \begin{equation}
% \left\{\begin{array}{rcl}
% \dot x&=&\dfrac{\partial H}{\partial p}=\dfrac p{m(t)}\\
% \dot p&=&-\dfrac{\partial H}{\partial x}=-m(t)\omega^2(t)\, x
% \end{array}\right.\,.\label{HeqTDHO}
% \end{equation}
% Its
% solutions are the integral curves for the $t$-dependent vector field 
% $$X(t)=\frac 1{ m(t)}\,p\,\pd{}x-m(t)\omega^2(t)\,x\pd{}p\,.
% $$
% If we consider the set of vector fields
% $$
% X_0=p\pd{}{x},\quad X_1=\frac 12\left(x\pd{}{x}-p\pd{}{p}\right),\quad X_2=-x\pd{}{p}\,,
% $$
% which close on a $\goth{sl}(2,\R)$ Lie algebra with commutation relations
% $$
% [X_0,X_1]=X_0,\quad [X_2,X_1]=-X_2,\quad [X_2,X_0]=-2X_1,
% $$
% then the $t$-dependent  vector field $X(t)$ 
% can be written as a linear combination
% $$
%  X(t)=m(t)\omega^2(t)\,X_2+\frac 1{m(t)}\,X_0\,, 
% $$
% i.e. it is a linear combination with $t$-dependent coefficients 
% $$
% X(t)=\sum_{\alpha=0}^2b_\alpha(t)X_\alpha,
% $$
%  with 
%  $$b_0(t)=\frac 1{m(t)}\,, \quad b_1(t)=0\,,\quad b_2(t)=m(t)\omega^2(t)\,.$$
% 
For instance, an harmonic oscillator with a damping term \cite{Cal41, {Ka48}} with equation of
motion 
$$\frac d{dt}(m_0\, \dot x)+ m_0\,\mu\,\dot x+k\, x=0\ ,\qquad k=m_0\omega^2,
$$
admits a Hamiltonian description with a $t$-dependent Hamiltonian
$$
H(t,x,p)=\frac{p^2}{2m_0}\exp(-\mu t)+\frac{1}{2}m_0\exp(\mu t)\omega^2x^2,
$$
i.e.  $m(t)$ in (\ref{TDHOH}) is  $m(t)=m_0\,\exp(\mu t)$. In this case 
the  equations (\ref{TDFHO2}) are
$$
\left\{\begin{array}{rcl}
\dot x&=&\dfrac 1{m_0}\exp(-\mu t)\,p,\\
\dot p&=&-m_0\, \exp(\mu t)\,\omega^2x,
\end{array}\right.
$$
and the $t$-dependent coefficients as a Lie system read
$$b_0(t)=\frac 1{m_0}\,\exp(-\mu t)\,, \quad b_1(t)=0\,,\quad  b_2(t)=m_0\omega^2\exp(\mu t)\,.
$$
Therefore, as $b_0(t)b_2(t)=\omega^2$, $b_1=0$ and 
$$\frac{\dot b_2}{b_2}-\frac{\dot b_0}{b_0}=2\mu\,,
$$
we see that (\ref{Kcond}) holds with $c_0=c_2=1, c_1=\mu/\omega$ and the function $D$ is
a constant $D=\omega$.  Hence, this
example reduces to the system
$$\frac{d}{dt}\matriz{c}{x'\\p'}=\matriz{cc}{\mu/2&\omega\\-\omega&-\mu/2}\matriz{c}{x'\\p'}$$
which can be easily integrated. Let $\bar\omega^2=({\mu^2}/{4})-\omega^2$, we get
{\small
\begin{equation*}
\left(\begin{array}{c}
x'(t)\\p'(t)
\end{array}\right)
=\left(\begin{array}{cc}
\ch(\bar\omega t)+\dfrac{\mu}{2\bar\omega}\sh(\bar\omega t)& \dfrac{\omega}{\bar\omega}\sh(\bar\omega t)\\
-\dfrac{\omega}{\bar\omega}\sh(\bar\omega t)& \ch(\bar\omega t)-\dfrac{\mu}{2\bar \omega}\sh(\bar\omega t)\\
\end{array}\right)
\left(\begin{array}{c}
x'(0)\\p'(0)
\end{array}\right)
\end{equation*}}
and, in terms of the initial variables, obtain
\begin{equation*}
x(t)=\frac{e^{-\mu t/2}}{\sqrt{m_0\omega}}\left(
\left(\ch(\bar\omega t)+\frac{\mu}{2\bar\omega}\sh(\bar\omega t)\right)\sqrt{m_0\omega}x_0+
\frac{\omega}{\bar\omega}\sh(\bar\omega t)\frac{p_0}{\sqrt{m_0\omega}} \right).
\end{equation*}

We can also study the TDFHO's described by the  Hamiltonian
\begin{equation*}
H(t,x,p)=\frac 12p^2+\frac 12F(t)\omega^2x^2\,,\quad F(t)>0,
\end{equation*}
where we assume, for simplicity, $m=1$. The $t$-dependent vector field $X$ is
$$X(t)=p\,\pd{}x-F(t)\omega^2x\pd{}p\,,
$$
which is a linear combination
$$
 X(t)=F(t)\omega^2X_2+X_0\,, 
$$
i.e. the $t$-dependent coefficients in (\ref{binX}) are
$$b_0(t)=1\,, \quad b_1(t)=0\,,\quad b_2(t)=F(t)\omega^2\,,
$$
and the condition for $F$ to satisfy (\ref{Kcond}) is
$$
\frac 12\, \frac{\dot F} F=c_1\,\omega\, \sqrt{F}.
$$
Therefore $F$ must be of the form 
\begin{equation*}
F(t)=\frac{1}{(L-c_1\omega t)^2}\,
\end{equation*}
and the Hamiltonian which can be exactly integrated is
\begin{equation*}
H(t,x,p)=\frac{p^2}{2}+\frac{1}{2}\frac{\omega^2}{(L-c_1\omega t)^2}x^2\,.
\end{equation*}
The corresponding Hamilton equations are
\begin{equation*}
\left\{\begin{aligned}
\dot x&=p,\\
\dot p&=-\frac{\omega^2}{(L-c_1\omega t)^2}\,x,
\end{aligned}\right.
\end{equation*}
and  the $t$-dependent change of variables  to be performed is
\begin{equation*}
\left\{\begin{aligned}
x'&=\sqrt{\frac{\omega}{L-c_1\omega t}}\,x,\\
p'&=\sqrt{\frac{L-c_1\omega t}{\omega}}\,p.
\end{aligned}\right.
\end{equation*}
Therefore
\begin{equation}\label{eq5}
\left\{\begin{aligned}
\frac{dx'}{dt}&=\frac{\omega}{L-c_1\omega t}\left(\frac{c_1}{2}x'+p'\right),\\
\frac{dp'}{dt}&=\frac{\omega}{L-c_1\omega t}\left(-x'-\frac{c_1}{2}p'\right).
\end{aligned}\right.
\end{equation}
Under the $t$-reparametrisation
\begin{equation*}
\tau(t)=\int^{t}_0\frac{\omega dt'}{L-c_1\omega t'}=\frac{1}{c_1}{\rm ln}
\left(\frac{K'}{L-c_1\omega t}\right),
\end{equation*}
the system (\ref{eq5}) becomes
\begin{equation*}
\left\{\begin{aligned}
\frac{dx'}{d\tau}&=\frac{c_1}{2}x'+p',\\
\frac{dp'}{d\tau}&=-x'-\frac{c_1}{2}p',
\end{aligned}\right.
\end{equation*}
which is equivalent to the transformed Caldirola--Kanai differential equation  through the change $\tau \mapsto \omega\, t$ and $c_1\mapsto \mu/\omega$. In any case, the solution is
\begin{equation*}
x'(\tau)=\left(\ch(\widetilde\omega\tau)+\frac{c_1}{2\widetilde\omega}\sh(\widetilde\omega\tau)\right)x'(0)+\frac{1}{\widetilde\omega}\sh(\widetilde\omega \tau) p '(0),
\end{equation*}
where $\widetilde\omega=\sqrt{\frac{c_1^2}{4}-1}$ and finally
{\footnotesize \begin{equation*}
x(\tau(t))=\sqrt{\frac{L-c_1\omega t}{\omega}}\left[\left(\ch(\widetilde\omega\tau(t))+\frac{c_1}{2\widetilde\omega}\sh(\widetilde\omega\tau(t))\right)x'(0)+\frac{1}{\widetilde\omega}\sh(\widetilde\omega \tau(t)) p '(0)\right].
\end{equation*}}

%  and as
% \begin{equation*}
% \tau(t)=\frac{1}{c_1}{\rm ln}\left(\frac{K'}{L-c_1\omega t}\right),
% \end{equation*}
% we get
% \begin{multline*}
% x(t)=\sqrt{\frac{L-c_1\omega t}{\omega}}\left[\ch\left(\frac{\tilde\omega}{c_1}{\rm ln}\left(\frac{K'}{L-c_1\omega t}\right)\right)\sqrt{\frac{\omega}{K'}}x(0)+\right.\\
% \frac{1}{\tilde\omega}\sh\left(\frac{\tilde\omega}{c_1}{\rm ln}\left(\frac{K'}{L-c_1\omega t}\right)\right)\left.\left(\sqrt{\frac{\omega}{K'}}x(0)+\sqrt{\frac{K'}{\omega}}p(0)\right)\right].
% \end{multline*}
\section{Another integrability condition.}
\qquad In this Section we analyse a new integrability condition that, as the
preceding ones,
 arises as a compatibility condition for a restricted case
of the system for the integral curves
of  (\ref{TRR4}). Nevertheless, this time the solution is restricted to
one-parameter sets of matrices of $SL(2,\R)$ that is not a group in general.

In this way, we deal with a family of transformations
\begin{equation}
\bar A_0(t)=\left(\begin{array}{cc}\frac{1}{V(t)}&0\\-u_1&
    V(t)\end{array}\right)\,,\qquad V(t)>0\,,\label{newfam}  
\end{equation}
where $u_1$ is a constant, i.e.  we want to relate the $t$-dependent vector field
\begin{equation*}
X(t)=X_0+F(t)\omega^2X_2,
\end{equation*}
characterised by the coefficients in (\ref{binX}) 
\begin{equation*}
b_0=1,\quad b_1=0,\quad b_2=F(t)\omega^2,
\end{equation*}
to an integrable one characterised by $b'_0,b'_1$ and $b'_2$,
or more explicitly, to the $t$-dependent vector field 
\begin{equation*}
X(t)=D(t)(c_0X_0+c_2X_2)\,,
\end{equation*}
i.e. $b'_0=Dc_0$, $b'_1=0$, and $b'_2=Dc_2$. Moreover, if $c_0\ne 0$ we can
reabsorb its value with a redefinition of $D$ and we
can assume $c_0=1$.

Under the action of  (\ref{newfam})  the original system becomes 
 the following system
\begin{equation*}
\left\{\begin{array}{rl}
b'_2&= V^2b_2+u_1Vb_1+u_1^2b_0-u_1\dot V,\\
b'_1&=b_1+2\dfrac {u_1}Vb_0-2\dfrac{\dot V}V,\\
b'_0&=\dfrac 1{V^2}b_0\,.\end{array}\right.
\end{equation*}

As $b_1=b'_1=0$ and $b_0=1$, the second  equation shows that $\dot V=u_1$,
i.e. $V(t)=u_1t+u_0$ with $u_0\in \R$. Moreover, using this condition on the
first equation together with $b_0=1$,  it becomes $b'_2= V^2b_2$. Then, as the third equation gives us the value of $D$ as $D=b'_0=1/V^2$,
we see that $b'_2=Dc_2=V^2F(t)\omega^2$. Therefore, $F$ must be 
 proportional to $(u_1t+u_0)^{-4}$, 
$$F(t)=\frac k{(u_1t+u_0)^{4}}\,, \qquad k=\frac{c_2}{\omega^2}\,.
$$
Let assume $k=1$ and thus $c_2=\omega^2$.

Then,  the $t$-dependent transformation $\bar A_0(t)$ 
performing  this reduction is 
\begin{equation*}
\left\{\begin{aligned}
x'&=\frac{x}{V(t)},\\
p'&=-u_1x+V(t)p.
\end{aligned}\right.
\end{equation*}
Under this transformation, the initial system becomes
\begin{equation*}
\left\{\begin{aligned}
\frac{dx'}{dt}&=\frac{p'}{V^2(t)},\\
\frac{dp'}{dt}&=-\frac{\omega^2x'}{V^2(t)}\,.
\end{aligned}\right.
\end{equation*}
Using           the $t$-reparametrisation
\begin{equation*}
\tau(t)=\int^t_0\frac{ dt'}{V^2(t')}=\frac{1}{u_1}\left(\frac{1}{u_0}-\frac{1}{ V(t)}\right),
\end{equation*}
we get the next autonomous linear system
\begin{equation*}
\left\{\begin{aligned}
\frac{dx'}{d\tau}&=p',\\
\frac{dp'}{d\tau}&=-\omega^2x',
\end{aligned}\right.
\end{equation*}
whose solution is
\begin{equation*}
\left(\begin{array}{c}x'(\tau)\\ p'(\tau)\end{array}\right)=
\left(\begin{array}{cc}\cos(\omega \tau)&\frac{\sin(\omega \tau)}{\omega}\\-\omega\sin(\omega \tau)&\cos(\omega \tau)\end{array}\right)
\left(\begin{array}{c}x'(0)\\ p'(0)\,,\end{array}\right).
\end{equation*}
Thus, we obtain that
\begin{equation*}
x(t)=V(t)\left(\cos(\omega\, \tau(t))\frac{x_0}{u_0}+\frac{1}{\omega}\sin(\omega\,\tau(t))(-u_1x_0+u_0p_0)\right)\,.
\end{equation*}
\section{Some other integrable systems.}
\qquad In this Section we show that the autonomisations of the transformed integrable systems obtained in latter Sections enable constructing $t$-dependent constants of the motion. Indeed, in previous cases, a TDFHO was transformed into a Lie system related to an equation in $SL(2,\R)$ 
\begin{equation*}
R_{A^{-1}*A}\dot A=-D(t)\left(c_0M_0+c_1M_1+c_2 M_2\right),
\end{equation*}
associated with a TDFHO determined by the  $t$-dependent vector field 
\begin{equation*}
X(t)=D(t)(c_0X_0+c_1X_1+c_2 X_2).
\end{equation*}
Any $t$-dependent first-integral $I(t)$ of this differential equation satisfies
\begin{equation*}
\frac{dI}{dt}=\pd{I}{t}+X(t) I=0.
\end{equation*}
Thus, the function $I$ is a first-integral for the vector field on $\R\times {\rm T}^*\R$ 
\begin{equation*}
\bar X(t) =c_0X_0(t)
  +c_1X_1(t)    +c_2 X_2(t)    +\frac{1}{D(t)}   \pd{}{t}.
\end{equation*}

 As $\R\times {\rm T}^*\R$ is a manifold with dimension three and the
 differential equation we are studying are determined by a distribution of
 dimension one, there are two independent
first-integrals.
%
%\begin{theorem} Any TDFHO which can be reduced by means of an integrability condition is superintegrable.
%\end{theorem}

Next, we analyse some integrable cases and their corresponding integrals.
\medskip

$\bullet$ Case $F(t)=(u_1t+u_0)^{-2}$:

\medskip

In this case we obtain that by Theorem 2 the  $t$-dependent vector field of the
initial TDFHO is transformed into the following one
\begin{equation*}
X(t)=\frac{\omega}{u_1t+u_0}\left(X_0    -\frac{u_1}{\omega}X_1    +X_2    \right)
\end{equation*}
and thus, using the method of characteristics, we obtain the following
first-integrals 
for this TDFHO:
\begin{equation*}
\begin{aligned}
I_1&=-\frac{u_1}{\omega}p'\, x' +x'^2+p'^2,\qquad
I_2&=\frac{(u_1+u_0 t)^{\omega/u_1}}{\left((\frac {u_1}{\omega}x'-2p')+2\bar\omega x'\right)^{\frac 1{\bar\omega}}},
\end{aligned}
\end{equation*}
with $\bar \omega =\pm\sqrt{\frac{u_1^2}{4\omega^2}-1}$.
\medskip

$\bullet$ Case $F(t)=(u_1t+u_0)^{-4}$:

\medskip

In this case we see that  the  $t$-dependent vector field of the 
initial TDFHO is transformed into
\begin{equation*}
X(t)=\frac{1}{V^2(t)}\left(X_0+\omega^2X_2\right),
\end{equation*}
and thus, using the method of characteristics, we get the following first-integrals
 for the initial TDFHO
\begin{equation*}
\begin{aligned}
I_1&=\left(\frac{x\,\omega}{V(t)}\right)^2+\left(V(t)p-u_1x\right)^2,\\
I_2&=\arcsin\left(\frac{x\,\omega}{V(t)\sqrt{I_1}}\right)+ \frac{\omega}{u_1V(t)}\,.
\end{aligned}
\end{equation*}
As we have two $t$-dependent constants of the motion in the space $\R\times {\rm T}^*\R$ and the solutions in this space are of the form $(t,x(t),p(t))$ we can obtain the solutions for our initial system.
\section{Conclusions and Outlook.}

\qquad  The first concern of this article has been to present a discussion on
integrability conditions from the viewpoint of Lie systems. Such a
discussion has been used to apply some previous results of
 the theory of integrability of Riccati equations to investigate $t$-dependent
 frequency harmonic oscillators. In this way, we have illustrated the use of
 our theory foe a well-known physical model obtaining particular integrable
 cases and
 constants of the motion. 

The procedure here developed can be straightforwardly used to study any other Lie system. A detailed lecture of the paper clarify that in the particular case of Riccati equations and TDFHO's the theory of Lie systems enable us:

\begin{itemize}
\item To reduce the problem of integrability for such Lie systems to the problem of integrability of equations in $SL(2,\mathbb{R})$.
\item  To show that integrability conditions for Lie systems in
  $SL(2,\mathbb{R})$ can be applied to any other Lie system associated with
  such equations, i.e. Milne-Pinney equations and Ermakov systems.
\item To note that integrability conditions of Lie systems are closely related to some kind of matrix Riccati equations. Moreover, the study of solutions of such equations with algebraic conditions describe integrability conditions of Lie systems.
\end{itemize}

Additionally, recent results on the theory of Lie systems \cite{CRL07c} allow
us to use the procedure here developed  to study at the same footing problems
in Quantum Mechanics \cite{CL09} and partial differential equations. It can
also be shown that PDE's consisting on matrix Riccati equations can be used to
analyse 
integrability conditions for PDE Lie systems. 

In a recent paper \cite{CGL09} we have proposed a generalisation of the concept of Lie
system 
which share many of its characteristic and can be used for a more general class
of systems as for instance dissipative Milne--Pinney equations \cite{CL09b}, Emden and Abel
equations.

\section*{Acknowledgements.}

 Partial financial support by research projects MTM2006-10531 and E24/1 (DGA)
 are acknowledged. JdL also acknowledge
 a F.P.U. grant from  Ministerio de Educaci\'on y Ciencia.

\end{document}